\documentclass[reprint,aps,prl,superscriptaddress,longbibliography]{revtex4-1}
\usepackage[utf8]{inputenc}
\usepackage{color}
\usepackage{times}
\usepackage{amsmath}
\usepackage{amssymb}
\usepackage{stmaryrd}
\usepackage{makecell}
\usepackage[caption=false]{subfig}
\usepackage{graphicx}
\usepackage{tikz-cd}
\usepackage[unicode=true,
 bookmarks=true,bookmarksnumbered=false,bookmarksopen=false,
 breaklinks=false,pdfborder={0 0 1},backref=false,colorlinks=true]
 {hyperref}
\hypersetup{
 linkcolor=magenta, urlcolor=blue, citecolor=blue, pdfstartview={FitH}, hyperfootnotes=true, unicode=true}
\makeatletter
\@ifundefined{textcolor}{}
{%
 \definecolor{BLACK}{gray}{0}
 \definecolor{WHITE}{gray}{1}
 \definecolor{RED}{rgb}{1,0,0}
 \definecolor{GREEN}{rgb}{0,1,0}
 \definecolor{BLUE}{rgb}{0,0,1}
 \definecolor{CYAN}{cmyk}{1,0,0,0}
 \definecolor{MAGENTA}{cmyk}{0,1,0,0}
 \definecolor{YELLOW}{cmyk}{0,0,1,0}
}

\newcommand{\nn}{\nonumber}

\def\be{\begin{equation}}
\def\ee{\end{equation}}
\def\bc{\begin{center}}
\def\ec{\end{center}}

\def\nn{\nonumber}

\def\r2{{\sqrt{2}}}

\def\Tr{{\rm Tr}}

\def\bea{\begin{eqnarray}}
\def\eea{\end{eqnarray}}

\definecolor{burntorange}{rgb}{0.8, 0.33, 0.0}

\newcommand{\doublewidetilde}[1]{{%
  \mathpalette\double@widetilde{#1}%
}}
\newcommand{\double@widetilde}[2]{%
  \sbox\z@{$\m@th#1\widetilde{#2}$}%
  \ht\z@=.9\ht\z@
  \widetilde{\box\z@}%
}

\usepackage{amsfonts}\usepackage{tabularx}\usepackage{dcolumn}\usepackage{bm}\usepackage{graphicx}\usepackage{epstopdf}

\hypersetup{urlcolor=blue}

\usepackage{tensor}
\usepackage{braket}

\usepackage[capitalise,compress]{cleveref}
\crefname{section}{Sec.}{Secs.}
\Crefname{section}{Section}{Sections}
\crefrangelabelformat{equation}{\textup{(#3#1#4)}--\textup{(#5#2#6)}}

\makeatother

\usepackage{xcolor}

\newcommand{\mycirc}[1][black]{\Large\textcolor{#1}{\ensuremath\bullet}}

\definecolor{darkgreen}{rgb}{0.0, 0.6, 0.13}

\usepackage{amsfonts}\usepackage{tabularx}\usepackage{dcolumn}\usepackage{bm}\usepackage{graphicx}\usepackage{epstopdf}

\hypersetup{urlcolor=blue}

\makeatother
\begin{document}

\title{Decoding conformal field theories: from supervised to unsupervised learning}

\author{En-Jui Kuo}
\affiliation{Department of Physics, University of Maryland, College Park, MD 20742, USA}
\affiliation{Joint Quantum Institute, NIST/University of Maryland, College Park, MD 20742, USA}

\author{Alireza Seif}
\affiliation{Department of Physics, University of Maryland, College Park, MD 20742, USA}
\affiliation{Joint Quantum Institute, NIST/University of Maryland, College Park, MD 20742, USA}
\affiliation{Pritzker School of Molecular Engineering, University of Chicago, Chicago, IL 60637}
\author{Rex Lundgren}
\affiliation{Joint Quantum Institute, NIST/University of Maryland, College Park, MD 20742, USA}
\affiliation{Joint Center for Quantum Information and Computer Science, NIST/University of Maryland, College Park, MD 20742, USA}

\author{Seth Whitsitt}
\affiliation{Joint Quantum Institute, NIST/University of Maryland, College Park, MD 20742, USA}
\affiliation{Joint Center for Quantum Information and Computer Science, NIST/University of Maryland, College Park, MD 20742, USA}

\author{Mohammad Hafezi}
\affiliation{Department of Physics, University of Maryland, College Park, MD 20742, USA}
\affiliation{Joint Quantum Institute, NIST/University of Maryland, College Park, MD 20742, USA}
\affiliation{Department of Electrical and Computer Engineering, University of Maryland, College Park, Maryland 20742, USA}

\begin{abstract}
We use machine learning to classify rational two-dimensional conformal field theories. We first use the energy spectra of these minimal models to train a supervised learning algorithm. We find that the machine is able to correctly predict the nature and the value of critical points of several strongly correlated spin models using only their energy spectra. This is in contrast to previous works that use machine learning to classify different phases of matter, but do not reveal the nature of the critical point between phases. Given that the ground-state entanglement Hamiltonian of certain topological phases of matter are also described by conformal field theories, we use supervised learning on R\'{e}yni entropies, and find that the machine is able to identify which conformal field theory describes the entanglement Hamiltonian with only the lowest few R\'{e}yni entropies to a high degree of accuracy. Finally, using autoencoders, an unsupervised learning algorithm, we find a hidden variable that has a direct correlation with the central charge and discuss prospects for using machine learning to investigate other conformal field theories, including higher-dimensional ones. Our results highlight that machine learning can be used to find and characterize critical points and also hint at the intriguing possibility to use machine learning to learn about more complex conformal field theories.
\end{abstract}


\maketitle

{\it Introduction.---} 
Conformal field theories (CFT), which are quantum field theories with conformal invariance, appear in many areas of physics including condensed matter, statistical physics, and string theory \cite{francesco2012conformal,ginsparg1988applied}. 
This procedure turns out to be especially powerful in two spacetime dimensions (one spatial dimensional and one temporal dimension), where the conformal group is infinite-dimensional, and certain two-dimensional CFTs may be classified by a finite number of primary fields \cite{francesco2012conformal,ginsparg1988applied}. These CFTs, which are realized in a number of physically relevant systems, including the low-energy theory of the quantum critical point of the transverse-field Ising model \citep{3disingbootstrap}, the edge states (along with the ground-state entanglement Hamiltonian) of fractional quantum Hall systems \cite{moore1991nonabelions,PhysRevLett.101.010504}, and the Polyakov action describing the world sheet in string theory \citep{polchinski1998string}, are important for being rare examples of analytically tractable strongly-interacting quantum field theories.

Therefore, given some data of a quantum system, it is important to identify whether that system is described by a CFT. This data, obtained from either experimental measurements or numerical simulations, could be the few lowest energy levels of a given Hamiltonian or R\'{e}nyi entanglement entropies, which can be measured by probing multiple copies of the system's state \cite{PhysRevB.85.035409,daley2012measuring, islam2015measuring, pichler2016measurement,wang2020calculating}. In particular, it is interesting to ask if this information can be used to detect whether a system is at a critical point, and what kind of CFT describes it the best. To address this question, we turn to machine learning.


Machine learning has been increasingly used to study a wide range of problems in different areas of physics over the past few years \cite{carleo2019machine}. Notable examples include: classifying phases of matter \cite{carrasquilla2017machine,wetzel2017unsupervised,van2017learning}, studying non-equilibrium dynamics of physical systems \cite{van2018learning,schindler2017probing,seif2019machine}, studying the string theory landscape \cite{carifio2017machine} and AdS/CFT correspondence \cite{PhysRevD.98.046019},  simulating dynamics of quantum systems\cite{carleo2017solving}, quantum state tomography \cite{torlai2018neural,carrasquilla2019reconstructing}, and augmenting capabilities of quantum devices \cite{seif2018machine,torlai2019integrating}.

\begin{figure}[t]
	\centering
	\includegraphics[width=\columnwidth]{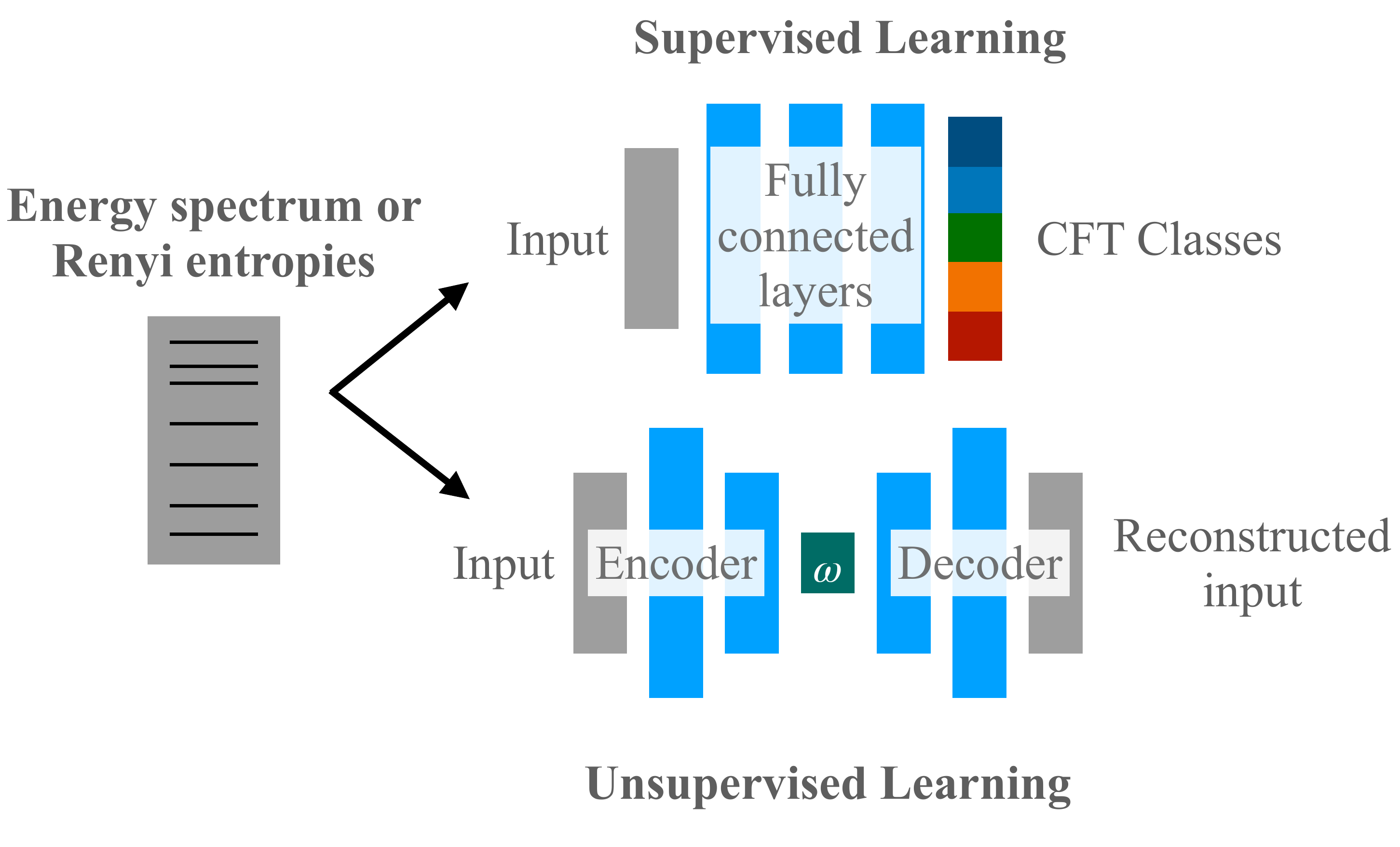}
	\caption{Schematic illustration of the machine-learning algorithms to identify different CFTs. The pre-processed energy spectrum or Renyi entropies are stored into a vector that serves as an input to the algorithms. We consider two scenarios: supervised learning, and unsupervised learning. In the former, labels of the CFT class are provided to a neural network classifier, which predicts the CFT theory that describes the given CFT data. In the latter, we use an autoencoder neural network, which learns an efficient representation of the energy spectra. The first half of the network acts as an encoder that maps the input to a single scalar variable $\omega$, and the second half decodes $\omega$ and reconstructs the original input. We find that ${\omega}$ is directly correlated to the central charge.}
	\label{fig:show}
\end{figure}

In this work, we use both unsupervised and supervised machine learning to investigate various two-dimensional CFTs, as sketched in Fig.\ref{fig:show}. For supervised learning, we use a deep neural network. Our first training data set is the lowest energy levels of exactly-solvable two-dimensional CFTs. 
The chosen CFT models include the well-known Ising critical point and $SU(2)_k$ anyonic chain parafermonic model (see Table~\ref{TAB:cftlist1} for a full list). We then ask the machine to locate and predict the nature of critical points of quantum spin-chains to high accuracy. By looking at the confidence of the network, we are able to correctly identify the value of the critical point. Remarkably, our approach requires a single system size, whereas common methods (such as entanglement scaling \cite{Calabrese_2009}) require finite-size scaling. Given that the entanglement spectra (ES) of various topological phases of matter are also described by CFTs \cite{PhysRevLett.101.010504,PhysRevA.86.032326}, we train our network with the lowest few R\'{e}yni entropies. While the relationship between R\'{e}yni entropies and ES is non-linear and requires all R\'{e}yni entropies to solve for the ES exactly, we are able to extract the CFT that describes the ES of two different spin-ladder systems to high accuracy with only having access to the lowest few R\'{e}yni entropies. Finally, we use the autoencoder algorithm \cite{goodfellow2016deep}, i.e. an unsupervised learning algorithm, and we find that the value of the hidden variable is directly related to the central charge. This gives us a hint that the machine can detect the complexity of CFTs.

{\it Machine learning and CFT basics.---}We first review the CFT knowledge needed to generate our training data (see Ref.~\cite{francesco2012conformal} for a detailed review of CFTs and Ref.~\cite{goodfellow2016deep} for machine learning), which is taken to be the lowest twenty energy levels of a finite-size model. We take our system to have periodic boundary conditions, although our approach can be readily generalized to include other boundary conditions. In this work, we restrict ourselves to rational CFTs (RCFTs), which only contain a finite number of primary fields, and we furthermore focus on CFTs with field content such that they are modular invariant (see the Supplementary Material \ref{app:2dcft} for definitions and details).
Our methods may easily be applied to CFTs with non-modular invariant field content.

The discrete energy levels (in units of $2\pi/L$, where $L$ is the length of the system, and $\hbar=1$) of a generic finite one-dimensional model which flows to a CFT is given by \cite{francesco2012conformal}
\be \label{energy}
E=E_1L + E_0 + \frac{2\pi v}{L}\left(-\frac{c}{12}+h_L+h_R \right),
\ee
where $E_{1}$, $E_0$, and $v$ are non-universal constants, and $c$ is the central charge of the CFT. We are also omitting subleading dependence on $L$ due to corrections to the scaling limit. Here, $h_L= h^{(0)}_L +m_L$ and $h_R= h^{(0)}_R +m_R$, where $h^{(0)}_L, h^{(0)}_R$ correspond to scaling dimension of the primary fields and $m_L$ and $m_R$ are non-negative integers describing the descendant fields. 

As a definite example, we now discuss the structure of the primary descendant fields for the critical Ising model, the simplest non-trivial CFT and an example of a Virasoro minimal model. With modular invariance imposed, there are three primary fields for this model, $h^{(0)}_{L,R}=0,\frac{1}{16}$ and $\frac{1}{2}$. The number of descendant fields can be calculated by expanding the so-called character function, as reviewed in the Supplementary Material \label{app:2dcft}. Upon doing so, one finds the lowest ten energy levels of this CFT (with the ground state energy set to zero) are $\frac{2\pi v}{L}\times\{0, \frac{1}{8}, 1, \frac{9}{8}, \frac{9}{8}, 2, 2,2,2,\frac{17}{8}\}$. The energy spectrum of some of the other models we consider are discussed in the Supplemental Material, and the list of all CFTs we consider for unsupervised learning is given in Table \ref{TAB:cftlist1}. We stress that this is by no means a complete list of RCFTs, as there are actually an infinite number of them.

We use a neural network to classify the input CFT spectra into their corresponding CFT classes (see Fig.~\ref{fig:show}). Specifically, we use the following neural network architecture with the input and output being the first 15 energy levels and their corresponding CFT class labels from the 13 CFT classes in Table \ref{TAB:cftlist1}:
\begin{tikzcd}
\mathbb{R}^{15} \arrow[r,"\text{linear}", rightarrow] & \mathbb{R}^{5}  \arrow[r,"\text{relu}", rightarrow] & \mathbb{R}^{13} \arrow[r,"\text{softmax}", rightarrow] & \mathbb{R}^{13}
\end{tikzcd}. The layers are represented by their domain and the expressions above the arrows indicate the activation functions of each layer. 
To train the network we take samples of the energy spectra of different CFT classes and add a noise term drawn randomly from the uniform distribution in $(-\epsilon, \epsilon).$ This is physically motivated by the existence of experimental measurement errors or subleading corrections to Eq.~\eqref{energy}. It also serves as a form of data augmentation than can prevent overfitting~\cite{goodfellow2016deep}. We also preprocess the input such that the ground state energy is set to zero and the other energies are rescaled so that largest energy level is 1. This removes the contributions of the non-universal constants in the input data (see Supp. Mat.). We then optimize the categorical cross entropy over 3000 samples for each class with $\epsilon=0.1$. The optimization is performed using the Adam optimizer with hyperparameters given in Ref.~\cite{kingma2014adam} in $2000$ epochs with the batch size set to 128.


\definecolor{0col}{RGB}{0.0, 0.0, 127}
\definecolor{1col}{RGB}{0.0, 0.0, 255.0}
\definecolor{2col}{RGB}{0, 99, 255}
\definecolor{3col}{RGB}{0, 212, 255}
\definecolor{4col}{RGB}{169, 255, 77}
\definecolor{5col}{RGB}{205, 255, 0}
\definecolor{6col}{RGB}{255, 124, 0}
\definecolor{7col}{RGB}{255, 19, 0}
\definecolor{8col}{RGB}{127, 0, 0}

\begin{table}[]
	\begin{tabular}{|c| c| c|}\hline
		Model                & Class & Central charge  \\\hline
		$(A_3, A_2)$ - Ising  & $0$ \mycirc[0col] & $1/2$ \\\hline
		$(A_4, A_3)$ - Tricritical & $1$ \mycirc[1col]  &$7/10$   \\\hline
		$(D_4,A_4)$  & $2$ \mycirc[2col] & $4/5$ \\\hline
		$(A_6, D_4)$ & $3$ \mycirc[3col] & $6/7$  \\\hline
		$\mathbb{Z}_4$ parafermion & 4 \mycirc[4col]  &$1$  \\ \hline
		$\mathbb{Z}_5$ parafermion & 5 \mycirc[5col] & $8/7$ \\ \hline
		$\mathbb{Z}_6$ parafermion & 6 \mycirc[6col] & $5/4$ \\ \hline
		$N=1$ SCFT, $k=5$ &  7 \mycirc[7col] & $81/70$   \\\hline
		$N=1$ SCFT, $k=7$  &  8 \mycirc[8col]  & $55/42$   \\\hline
		$(A_{k+1}, A_k), k=5-8$ & $9-12 $  & $1-6/(k(k+1))$ \\\hline
	\end{tabular}
\caption{List of all conformal field theories, i.e. classes, we include in our supervised training. We  use color dots to indicate the CFT shown in Figs. \ref{fig:ising} and \ref{fig:H3}.} 
\label{TAB:cftlist1}
\end{table}

\begin{figure}
	\centering
	\includegraphics[width=\columnwidth]{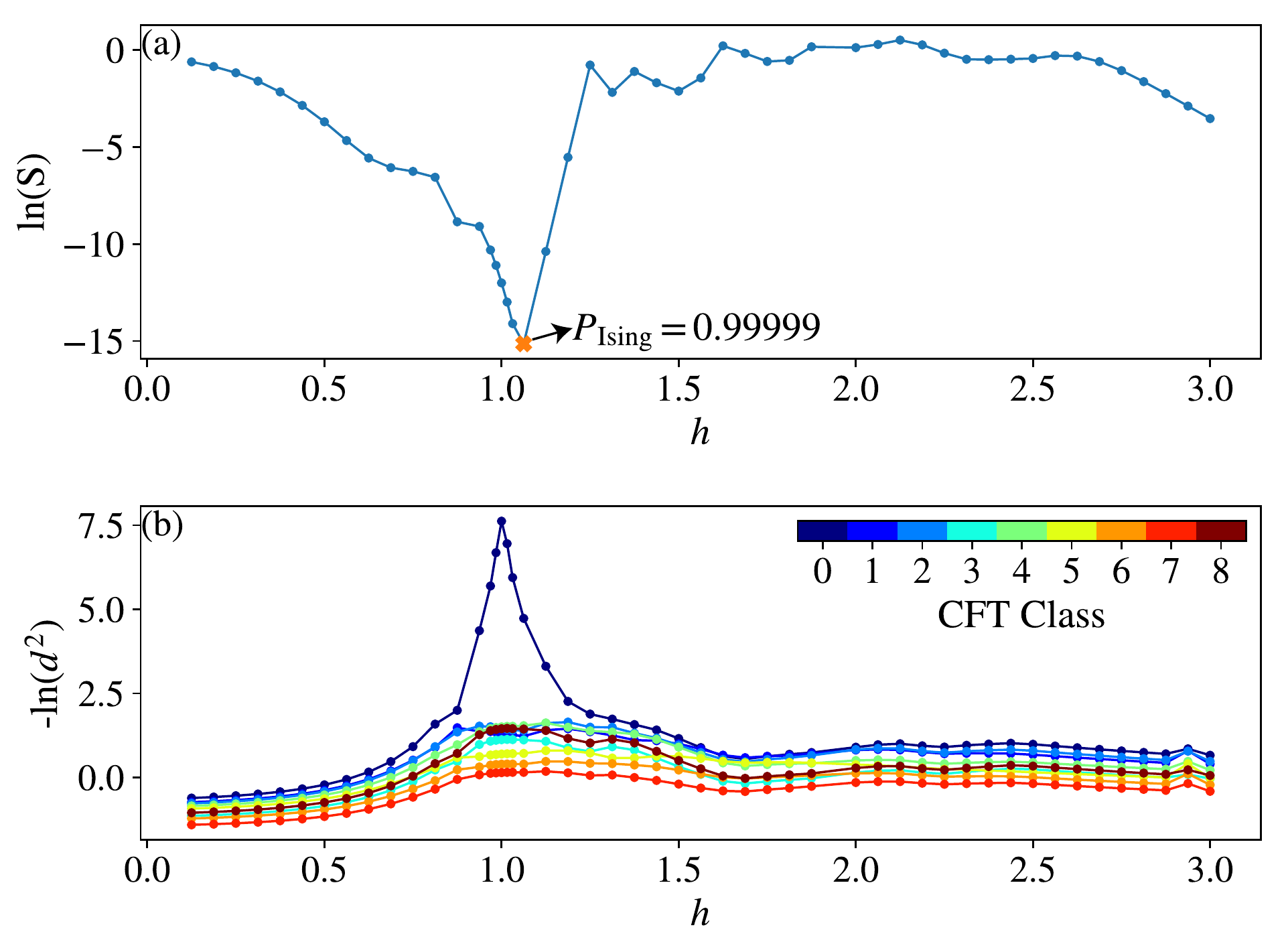}
	\caption{Predictions of our supervised machine learning approaches on the spectrum data transverse field Ising model as a function of the transverse field $h$. (a) The confidence, i.e entropy, $S$, of the neural network. (b) The Gaussian distance, $d_2$, between the energy spectrum of $H_{\mathrm{ISING}}$ and the energy spectrum of the first 9 CFTs in Table~\ref{TAB:cftlist1}. The numbers in the legend refer to the CFT class.}
	\label{fig:ising}
\end{figure}

\begin{figure}
	\centering
	\includegraphics[width=\columnwidth]{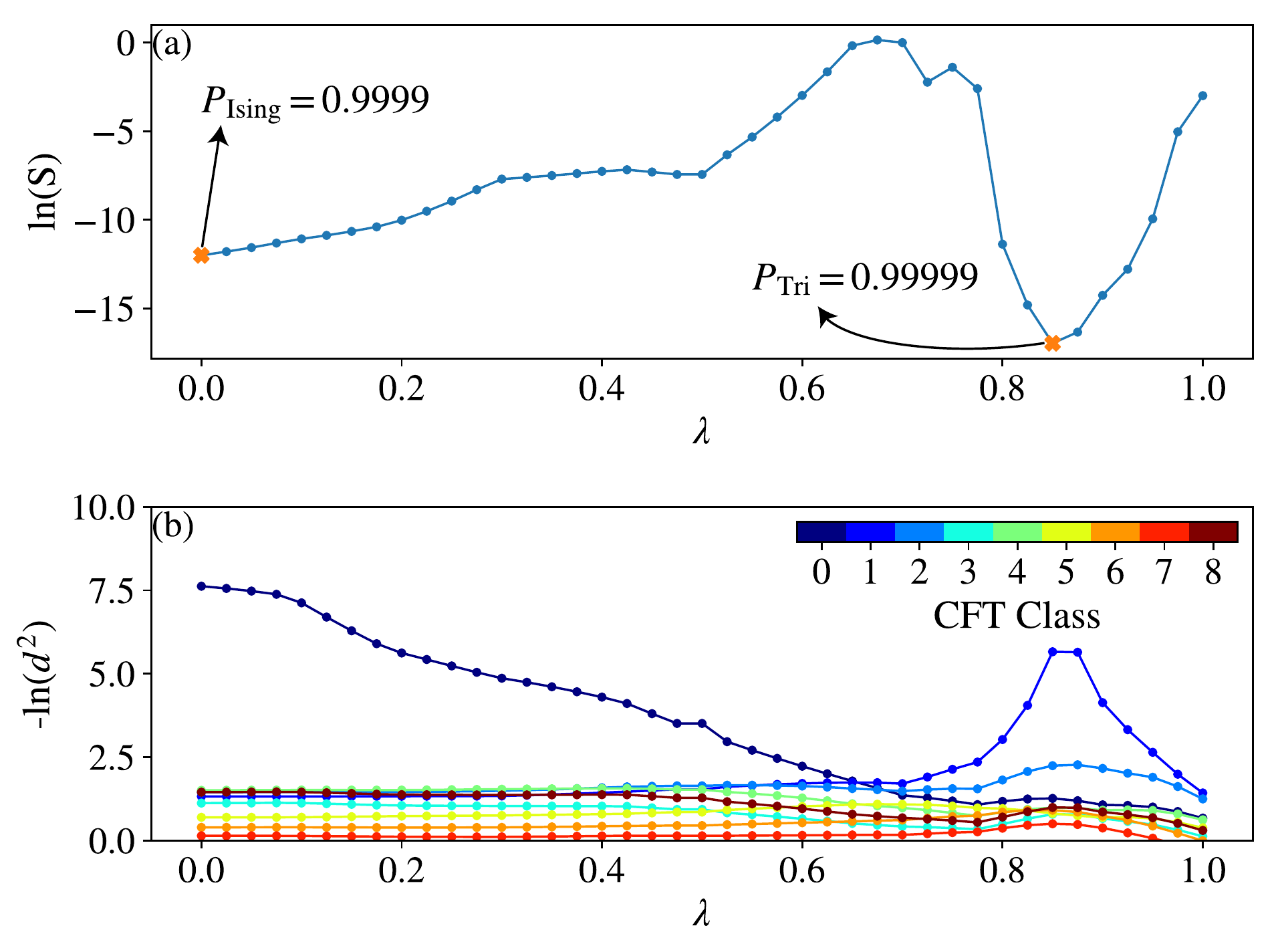}
	\caption{Predictions of our supervised machine learning approaches on a spin model composed of the transverse field Ising Hamiltonian with $h=1$ (see Fig.~\ref{fig:ising}) with added three-body interaction terms that are controlled by a parameter $\lambda$. (a) The entropy of probabilities, $S$, of the neural network. (b) The Gaussian distance, $d_2$, between the energy spectrum of $H_{T}$ and the energy spectrum of  the first 9 CFTs in Table~\ref{TAB:cftlist1}. The numbers in the legend refer to the CFT class.}
	\label{fig:H3}
\end{figure}

\paragraph{}

{\it Critical spin-chains.---} We now use the network trained on ideal CFT energy spectra with added noise, to make predictions for two physical models. Specifically, we feed the machine energy spectra from two many-body quantum spin models obtained using exact diagonalization. By analyzing the output of the machine, we are able to predict the location of the critical point and the type of CFT that describes it. 

We first consider the transverse-field Ising model, $H_{\mathrm{I}}(h)=-\sum_{i=1}^L \sigma^{z}_i \sigma^{z}_{i+1}-h\sum_{i
=1}^L \sigma^{x}_{i}$. Here, $\sigma_{\alpha}^i$ are the Pauli matrices on site $i$ and $h$ is a global magnetic field. For $h=1$, 
the low-energy theory of $H_{\mathrm{I}}$ is a CFT whose central charge is $c=1/2$ (minimal model $(A_3, A_2)$). We perform exact diagonalization (for $L=22$) for different $h$ and feed the energy spectrum into the network. In Fig.~\ref{fig:ising}a, we observe that the entropy $S$ of the output layer,  $S=-\sum_{i} p_i \ln p_i$, where $p_i$ corresponds to the $i$ th value of the output softmax layer, is minimal at $h=1$ and has a large probability of being described by minimal model $(A_3, A_2)$. This indicates that the network has not only correctly predicted the location of the critical point for this model, but also the nature of the critical point.

Before moving on to the next model, we discuss another quantitative approach to identifying CFTs from energy spectra. We consider a clustering algorithm, using the Gaussian kernel of the euclidean distance, i.e.,  $d_2(x,c_m)=e^{-||x-c_m||_2^2}$, where $x$ is the input spectra and $c_m$ is the center of the $m$'th cluster~\cite{goodfellow2016deep}. In our work, the ideal center of clusters is known from the CFT theory. Therefore, given an energy spectra, we calculate and compare the kernel on the rescaled input and the cluster centers of each CFT. In Fig. \ref{fig:ising}b, we plot $d_2$ for the Ising model and various CFTs. We observe that $d_2$ is peaked around the critical point for only the Ising model (Fig~\ref{fig:ising}b), similar to the neural network approach (Fig~\ref{fig:ising}a). However, we believe the neural network approach will be more reliable as it does not rely on a single energy spectrum. 

We now move to a more complicated model, which has two critical points described by different CFTs and both two-body and three-body interactions. The Hamiltonian of this model, originally introduced in Ref.~\cite{o2018lattice}, is given by $H_T=2H_{\mathrm{I}}(1)+\lambda H_3$, where $H_3 = \sum_{j}\sigma^{x}_{j}\sigma^{z}_{j+1}\sigma^{z}_{j+2}+\sigma^{z}_{j}\sigma^{z}_{j+1}\sigma^{x}_{j+2}$. For $\lambda=0$, this model is described by minimal model $(A_3, A_2)$ as discussed above. When $\lambda\approx0.856$, the low-energy theory of this model is described by a different minimal model, $(A_4, A_3)$. Again, we feed the network the many-body energy spectrum for various $\lambda$. In Fig. \ref{fig:H3}a we observe that the machine is correctly able to identify the location and underlying CFTs of the two critical points with high accuracy. Similar results are seen for the Gaussian kernel method (Fig.~\ref{fig:H3}b).

{\it R\'{e}nyi Entropies.---}
We now consider training with R\'{e}yni entropies. This is motivated by the fact that the (bipartite real-space) entanglement Hamiltonian, $H_e$, of two-dimensional topological phases is often described by (either chiral or non-chiral) one-dimensional CFTs \cite{PhysRevLett.101.010504,PhysRevA.86.032326}. 
Unfortunately, it is hard to experimentally measure the eigenvalues of $H_e$, i.e. the ES (although there are various theoretical proposals on how to do so \cite{PhysRevLett.117.010603}). Instead, one typically measures the R\'{e}yni entropy by preparing multiple copies of the state and interfering them \cite{islam2015measuring}. Furthermore, one can calculate $S_n$ with quantum Monte Carlo, making the calculation of entanglement more manageable for larger systems \cite{PhysRevB.89.195147, wang2020calculating}. We will demonstrate that, given a critical $H_e$ \footnote{When $H_e$ is far from criticality, our neural-network occasionally incorrectly predicts the sample data to be in one CFT class with very large probability. This is in sharp contrast to training with energy spectra. It would be interesting to see if this a fundamental problem with R\'{e}yni entropies or can be fixed by using more sophisticated techniques such as adversarial training.}, one can train neural-networks with R\'{e}yni entropies to correctly identify the underlying CFT.

The $n$-th R\'{e}yni entropy is defined as $S_{n}=\frac{1}{1-n}\ln\Tr\rho_A^n$, where $\rho_A$ is the reduced density matrix and $n$ is some positive integer not equal to one. The ES (eigenvalues of $-\ln(\rho_A)$) can be obtained with knowledge of $S_n$ for all $n$. In practice one can obtain an estimate of ES with only a finite number of $S_n$ \cite{PhysRevB.89.195147, PhysRevB.85.035409}. In these approaches, the ES is obtained from the roots of a polynomial equation, whose coefficients are related to R\'{e}yni entropies through Newton's identities. However, root-finding algorithms are sensitive to errors in the coefficients, making such schemes unstable in the presence of errors in $S_{n}$ measurements \cite{wilkinson1959evaluation}. We approach this problem using machine-learning.

If $H_e$ is a CFT, the $i$th ES level is given by $\epsilon_i=\epsilon_0+\epsilon_1L+\frac{2\pi v}{L}n_i$, where $n_i$ is the universal part of the spectrum (see Eq.~\eqref{energy}). We remind the reader $n_i$ is different set of numbers for each respective CFT. $S_n$, which is only a function of $\frac{v}{L}$ (for a given CFT), can then be written as
\begin{equation}
S_{n}=(1-n)^{-1}\ln [(\sum_{i=0}e^{-n \frac{2\pi v}{L} n_{i}})/(\sum_{i=0}e^{-\frac{2\pi v}{L}n_{i}})^{n}].
\label{reyni}
\end{equation}
We restrict the sum in Eq.~\eqref{reyni} to the lowest 100 ES levels. For training, we consider a finite range of $\frac{v}{L} \in (0.2, 10)$. This range is chosen not to include large (small) $\frac{v}{L}$ where excited state information is washed out (choice of cut-off plays an important role). Also, note that for larger $n$, $S_n$ becomes less dependent on the cut-off by definition. Thus, in the chosen range of $\frac{v}{L}$, the choice of cut-off, i.e. simply truncating the sum, has little effect on our results.

\begin{figure}
	\centering
	\includegraphics[width=\columnwidth]{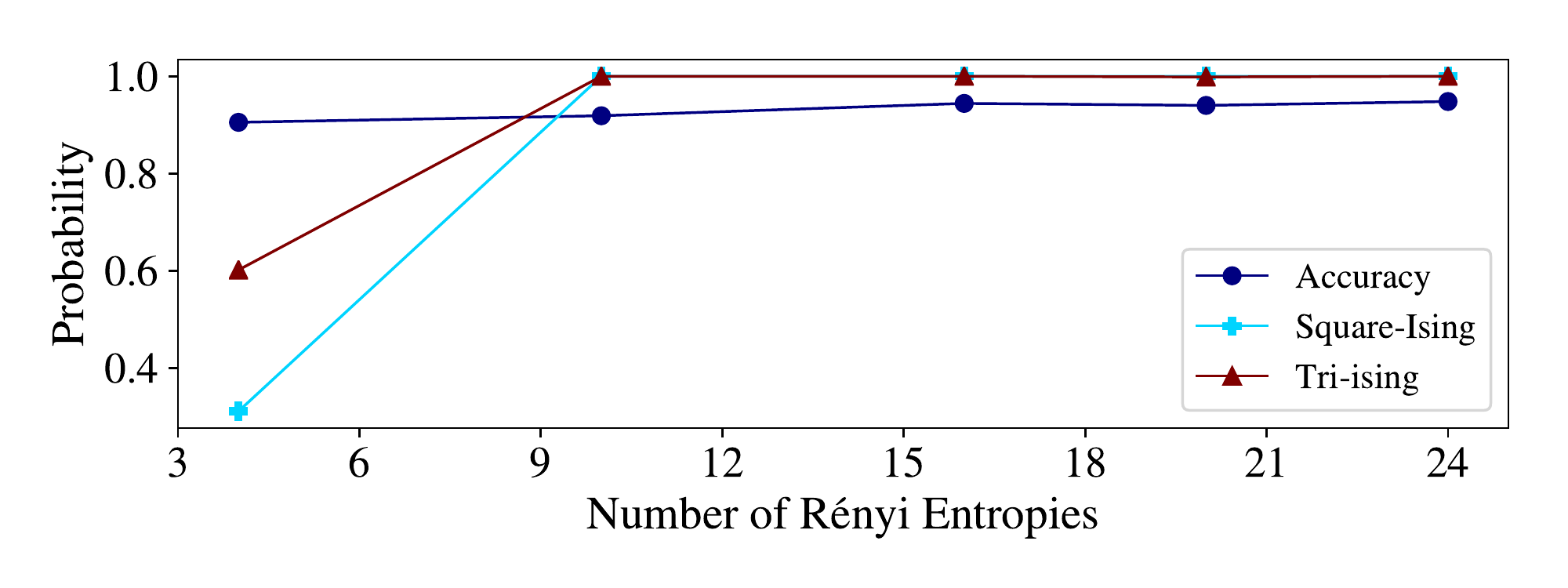}
	\caption{Probability of the neural network predicting the correct CFT of the entanglement Hamiltonian for the square (triangle) Ising model, depicted with square (diamond), as a function of number of R\'{e}yni entropies included in training samples.  The overall accuracy of the network is benchmarked using the test data,  shown as dark blue circles. }
	\label{fig:tri1}
\end{figure}

Instead of each sample being a vector of energy levels as in the previous section, it is a vector of R\'{e}yni entropies, $(S_2,S_3,\dots)$. Here, we include up to $28$ R\'{e}yni entropies, starting with $S_2$. We train our machine with $10000$ different samples for each CFT class (same classes used for energy spectrum training). We generate the data uniformly by randomly choosing $\frac{v}{L}$. We obtain a training accuracy of up to 94 \% depending on the number of $S_\alpha$ included. Generally, upon increasing the number of $S_\alpha$ included, the accuracy increases (see Fig.~\ref{fig:tri1}).  We now describe the training process. The network architecture is:
\begin{tikzcd}[row sep=tiny, column sep = small]
\mathbb{R}^{n} \arrow[r,"\text{relu}", rightarrow] & \mathbb{R}^{100}  \arrow[r,"\text{relu}", rightarrow] & \mathbb{R}^{20} \arrow[r,"\text{relu}", rightarrow] & \mathbb{R}^{15}  \arrow[r,"\text{relu}", rightarrow] & \mathbb{R}^{13} \arrow[r,"\text{softmax}", rightarrow] &[0.5em]  \mathbb{R}^{13}
\end{tikzcd}
where $n$ is the number of  R\'{e}yni entropies we use. Similar to the  classification of the energy spectra, we train the network by optimizing the cross-entropy using the Adam optimizer \cite{kingma2014adam}, this time with 500 epochs and batch size set to 128.

We now test our model on two exactly-solvable systems \cite{PhysRevA.86.032326}. The (unnormalized) ground state of these two models, which can be regarded as quantum spin ladders, can be written as $|\psi(z)\rangle =\sum_{\tau,\sigma}[T(z)]_{\tau,\sigma}|\tau\rangle|\sigma\rangle$. For the square ladder model, 
\begin{equation} \label{m1}
[T(z)]_{\tau,\sigma}=\prod_{i=1}^Lz^{(\sigma_i+\tau_i)/2}f(\sigma,\tau),
\end{equation}
and for the triangle ladder model,
\begin{equation} \label{m2}
[T(z)]_{\tau,\sigma}=\prod_{i=1}^Lz^{(\sigma_i+\tau_i)/2}f(\sigma,\tau)(1-\sigma_i\tau_{i+1}).
\end{equation}
Here, $\sigma_i$ and $\tau_i$ are either 0 or 1 and $f(\sigma,\tau)=(1-\sigma_i\tau_i)(1-\sigma_i\sigma_{i+1})(1-\tau_i\tau_{i+1})$. We then trace out one of the legs of the ladder. The spectrum of $\rho_A$ is identical to the spectrum of the following matrix, $M=\frac{1}{\Xi(z)}[T(z)]^TT(z)$, where $\Xi(z)$ ensures $\rho_A$ is properly normalized. One can interpret $T(z)$ as the transfer matrix of certain two-dimensional classical models with known critical points. Hence, if $T(z)$ is critical, $H_e$ will be critical \cite{PhysRevA.86.032326}. The critical point of the square (triangle) ladder is $z_c\approx 3.8 (11.1)$. The critical theories of the square and the triangle ladder models are described by minimal models $(A_3,A_2)$ and  $(D_4,A_4)$ respectively. We numerically calculate $S_n$ (from $M$) at $z_c$ for $L=18$ and use these numerical results as input into our trained neural-network. We remarkably find that the neural-network correctly predicts the CFT that describes $H_e$ for both models with high accuracy. As expected, this accuracy generally increases as one increases the number of $S_\alpha$ included in the training set (see Fig.~\ref{fig:tri1}).

\begin{figure}
	\centering
	\includegraphics[width=\columnwidth]{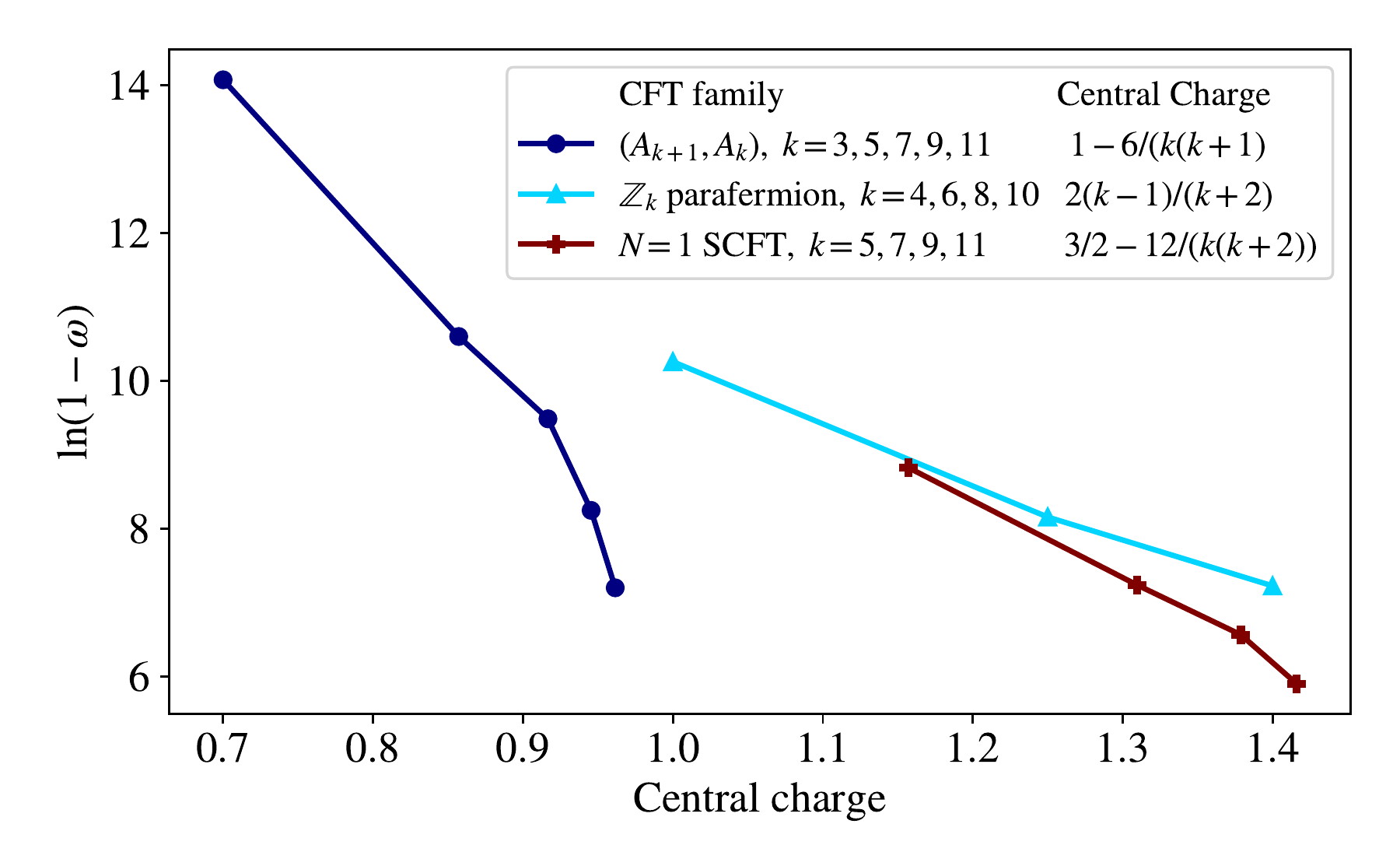}
	\caption{Unsupervised learning: The hidden variable as a function of the central charge for different CFT families (different colors and markers). We observe that within a family, the one-dimensional hidden variable $\omega$ is a monotonous function of the central charge.}
	\label{fig:auto}
\end{figure}

{\it Unsupervised Learning.---} We now turn to using unsupervised learning to explore two-dimensional CFTs.
Our data consists of three families of CFTs (see Fig.~\ref{fig:auto} for the list of CFTs used for unsupervised training).
We use autoencoders \cite{Goodfellow-et-al-2016} to find a compressed representation of the CFTs (see Fig.~\ref{fig:show}). Previously, autoencoders have been able to detect the order parameter, i.e. magnetization, in the Ising model \cite{wetzel2017unsupervised}. The autoencoder is comprised of an encoder function $\omega=f(\mathbf{x})$ and a decoder function $\mathbf{r}=g(\omega)$, where the hidden variable $\omega$ encodes a compressed representation of the input $\mathbf{x}$. The hidden variable is used by the decoder to find the reconstruction $\mathbf{r}$. By restricting the dimension of $\omega$, the network only approximately reconstructs the input, however, it learns the important features of the training data and encodes it in $\omega$. Each class has 100 examples which consists of the lowest 100 energies of the CFT (with same noise added as our energy based classification section.)

We train different autoencoders on a set of energies corresponding to different CFT classes by minimizing $C = \frac{1}{N_m}\sum_m \lvert|\mathbf{r}_m -\mathbf{x}_m|\rvert_2^2,$
where the sum is taken over $N_m$ examples in the training set. The $x_m$ is a input of the first layer and $r_m$ is a output of the last layer. $\mathbf{x}$ contains the first 100 energy levels. We use the following architecture for the autoencoder:
\begin{tikzcd}
\mathbb{R}^{100} \arrow[r,"\text{relu}", rightarrow] & \mathbb{R}^{5}  \arrow[r,"\text{sigmoid}", rightarrow] & \mathbb{R}^{h} \arrow[r,"\text{relu}", rightarrow] & \mathbb{R}^{5} \arrow[r,"\text{sigmoid}", rightarrow] &  \mathbb{R}^{100}, 
\end{tikzcd}
where $h$ is the dimension of the hidden variable $\omega$. We train the network by optmizing $C$, using the Adam optimizer with $2000$ epochs and batch sizes equals to $256$. We consider the simpliest case of $h=1$ and show the value of $\omega$ for different CFT spectra in Fig.~\ref{fig:auto}. We observe that within a single family of CFTs, the magnitude of the hidden variable has positive correlation with $k$, and hence the central charge. 
 
Finally, we note recent work used supervised machine learning to investigate CFT correlation functions and the emergence of conformal invariance \cite{2020arXiv200616114C}. Recently, it is been demonstrated that for the specific conformal field theory like Ising CFT, one can use an unsupervised learning method to classify them without dimensional reduction \cite{mendes2021unsupervised, mendes2021intrinsic}. However, the information of the critical point should be known in advance which is different from our work. Our work specifies conformal field theory by using the hidden variable when restricting to a single-family without knowing the critical point in advance.
In the future, it would be interesting to include correlation functions in our unsupervised training to see if could distinguish different families of CFTs.

{\it Discussion.---} We have investigated possible applications of machine learning for CFTs. By supervised training with two-dimensional CFT energy spectra and R\'{e}yni entropy, we find our network can identify the critical points of many-body models to high-accuracy. In the case of energy spectra, our network is also able predict the location of critical points. For the unsupervised learning part, we see that the latent variable increases with the central charge of the CFT (for a given CFT family) suggesting that machine learning can detect the complexity of CFTs.

There are several directions where our work can be readily extended. For example, the entanglement Hamiltonian of a CFT can always be written in terms of the boost operator \cite{BWtheorem,HislopLongo}, to which our methods may be applied. For critical one-dimensional systems, the entanglement spectra for particular entanglement cuts correspond with the spectra of boundary CFTs \cite{luchli2013operator,an2020entanglement, cardy2016entanglement}, which are also known \cite{cardyBCFT1,cardyBCFT2,Cho_2017}. We believe the methods developed in this Letter can be straight forwardly extended to these CFTs. Including more complicated two-dimensional CFTs in the energy spectrum training process (such as non-unitary CFTs \cite{bianchini2014entanglement,strominger2001ds,osborn2016c,martins1990connection}, floquet CFTs \cite{berdanier2017floquet,wen2018floquet,wen2018quantum}, CFTs with continuous scaling dimensions \cite{luchli2013operator}, and the product of two CFTs \cite{PhysRevB.94.081112}) would also be a worthwhile avenue to pursue.

Finally, it would be extremely interesting to use machine-learning techniques investigate higher-dimensional CFTs where the conformal group is no longer infinite dimensional and much less is understood about the structure of the energy spectrum. The most convenient geometry for exact diagonalization in 2$+$1D is the torus, where the low-energy energy spectrum is a universal fingerprint of the universality class described by the CFT. Recent studies have obtained the low-energy torus spectra for CFTs in several universality classes, including the Wilson-Fisher CFTs and related $\mathbb{Z}_2$ confinement transitions \cite{Schuler2016,Whitsitt2016,Whitsitt2017}, the fermionic chiral (or Gross-Neveu) fixed points \cite{schuler2019torus}, and QED3 \cite{Thomson2017}. In principle, the data obtained in these studies could be used as training data for a generalization of the present work to 2$+$1D, including the unsupervised learning portion. One could then imagine training with both standard order parameters \cite{carrasquilla2017machine,wetzel2017unsupervised,van2017learning} and energy spectra to establish phase diagrams and the nature of critical points in higher-dimensions.

\begin{acknowledgments}
We are grateful to M. Dalmonte, W. DeGottardi, and M. A. Rajabpour for useful discussions and S. Tanaka for sharing numerical data. E.J.K., A.S. and M.H. acknowledge support from AFOSR-MURI FA95501610323, U.S. Department of Energy, Quantum Systems Accelerator program and the Simons Foundation. A.S. is additionally supported by a Chicago Prize Postdoctoral Fellowship in Theoretical Quantum Science. R.L acknowledges support by the DoE BES QIS program (award No. DE-SC0019449), AFOSR, DoE ASCR Quantum Testbed Pathfinder program (award No. DE-SC0019040), NSF PFCQC program, NSF PFC at JQI, ARO MURI, and ARL CDQI. S.W. acknowledges support from the NIST NRC Postdoctoral Associateship award.
\end{acknowledgments}

\appendix

\section{Two-dimensional conformal field theories}
\label{app:2dcft}
\paragraph{}
In this section, we summarize the important aspects of the two-dimensional CFTs relevant to the results presented in this paper. Detailed discussions of two-dimensional CFTs can be found in Refs.~\cite{francesco2012conformal,ginsparg1988applied, qualls2015lectures}.

 The Virasoro minimal models are the complete set of unitary CFTs with a finite number of irreducible representations under the Virasoro algebra; however, if a CFT is also invariant under a larger symmetry group, it may be an RCFT by having a finite number of irreducible representations under the extended symmetry algebra. This is the case for parafermionic models and superconformal minimal models, which contain conserved parafermionic and fermionic currents respectively. 

The two-dimensional CFTs we consider may be specified by a central charge, $c$, and a finite set of holomorphic and anti-holomorphic fields, denoted by $\phi_{h_L}(z)$ and $\phi_{h_R}(\bar{z})$ respectively. Here, we use complex coordinates $z = x + i t$ and $\bar{z} = x - i t$ to parametrize the two-dimensional coordinates $(x,t)$. The numbers $(h_L,h_R)$, which are called the conformal dimensions of the associated primary fields, are real numbers which are generically independent. With this data, it is known that the finite-size energy spectrum of a two-dimensional CFT (in units of $2 \pi/L$) is given by Eq.~\eqref{energy}, where the lowest states correspond to primary fields, and the higher states are known as descendants. However, the degeneracy of the states corresponding to primary operators and their descendants can be nontrivial \cite{cardy1986operator}. 

We now review the degeneracy structure of the energy spectrum. In this work, we simply present the result for the partition function and refer the reader to Ref.~\cite{francesco2012conformal} for details. 
We consider an RCFT on a torus with complex-valued periods equal to $\omega_1$, $\omega_2$, and define the modular parameter of the torus as $\tau = \omega_2/\omega_1$.
Then we can write the partition function of an RCFT on the torus as \cite{francesco2012conformal}
\be
Z(\tau) = \sum_{h_L, h_R} M_{h_L,h_R} \chi_{h_L}(\tau)\chi_{h_R}(\bar{\tau}),
\label{eq:partdef}
\ee
where 
\be
\chi_{h_{L,R}}(\tau) = \sum_{n = 0}^{\infty} \mathrm{dim}(h_{L,R} + n) q^{h_{L,R} + n - c/24}
\ee
are the so-called characters associated with a given primary operator $\phi_{h_{L,R}}$.
Here, $M_{h_L,h_R}$ counts the number of occurrences of the primary $\phi_{h_L}(z) \times \phi_{h_R}(\bar{z})$ in the CFT, and we use the parametrization $q = e^{2 \pi i \tau}$. 

The reason for considering the partition functon on the torus is to demand that $Z(\tau)$ be left invariant under the modular transformations $\tau \rightarrow \tau + 1$ and $\tau \rightarrow -1/\tau$. This strongly constrains the structure of the spectrum investigated in the main body of the paper. It is believed that only modular invariant CFTs can be realized by a one-dimensional quantum lattice model, although non-modular invariant CFTs may arise as boundaries of two-dimensional lattice theories with bulk topological order \cite{Levin2013}. In the following we discuss the form of $\chi_{h_L}$ for the CFT families we are interested in.

\subsection{Virasoro minimal models}
In Virasoro minimal models, the central charge of the Virasoro algebra takes values of the type \cite{gils2013anyonic},
\be
c_{p,q}=1-6\frac{(p-q)^2}{pq},
\ee
where $p,q$ are coprime integers such that $ p,q\geq 2$. Then the allowed conformal dimensions of the (anti)holomorphic representations are
\be
h_{r,s}={\frac {(pr-qs)^{2}-(p-q)^{2}}{4pq}}\ ,\quad {\text{with}}\ r,s\in \mathbb {N} ^{*}\,
\ee
where
\be
{\displaystyle 1\leq r\leq q-1\quad ,\quad 1\leq s\leq p-1\ .}
\ee
The $\displaystyle (p,q)$ and ${\displaystyle (q,p)}$ models are the same.


From the previous discussion, we know that the allowed values of $(h_L,h_R)$ and their degeneracies can be inferred by the set of modular invariant partition functions on the torus.
The complete set of such partition functions has been entirely worked out for the unitary minimal models using the so-called ADE classification \citep{cappelli1987ade}.

As a definite example, we consider the Ising CFT ($c=1/2$), in which case there is only a single modular invariant choice of operators. If one expands the partition function in terms of the parameters $q=e^{2 \pi i \tau}$, $\bar{q}=e^{-2 \pi i \bar{\tau}}$, the full energy spectrum and its degeneracy can be read off from the coefficients and powers of the expansion. For the Ising CFT, the partition function turns out to be diagonal, meaning one only allows fields of the form $\phi_{h_L}(z) \times \phi_{h_R}(\bar{z})$ with $h_L = h_R$, and we can read off the spectrum from the degeneracy of $q$ alone. Just giving the $q$-dependence, and keeping terms up to $q^5$ and the level-$3$ descendants, the expansion is
\bea
& q^{\frac{1}{12} \left(1-\frac{6}{3 (4)}\right)}Z_{ising}=1+\sqrt[8]{q}+q+2 q^{9/8} +4 q^2\\ \nn 
& +3 q^{17/8}+5 q^3+O\left(q^{25/8}\right).
\eea
This expression should be read as follows: with energies measured with respect to the ground state and in units of $2 \pi/L$, we have unique states with $E=0, 1/8, 1$, a two-fold degenerate state with $E = 9/8$, a four-fold degenerate state with $E = 2$, a three-fold degenerate state with $E = 17/8$, etc. 
This explains how the energy level structure of the Ising model given in the main text was obtained.

\subsection{$\mathbb{Z}_{k}$ parafermion CFTs}
\label{appendix:para}
$\mathbb{Z}_{k}$ parafermion CFTs have a central charge given by $c=\frac{2(k-1)}{k+2}$ \cite{gils2013anyonic}. The conformal dimensions of the primary fields of these CFTs are
\be
h_{r,s} =\frac{r(r + 2)}
{4(k + 2)}-\frac{s^2}{4k},
\ee
where $r = 0, 1, . . . , k $ and $s =-r + 2, -l + 4, . . . , r$.
The associated character is~\cite{distler1990brs}
\be
\chi_{(r, s)}(\tau)=\eta(q)c_{s}^{r}(q),
\ee 
where $c_{s}^{r}(q)$ is given by
\bea
& c_{s}^{r}(q) = \sum_{l,m=0}^{\infty}(-1)^{l+m} q^{(k+r)lm+\frac{1}{2} (l+1) l+\frac{1}{2} (m+1) m}\times \nn \\ 
& \frac{q^{-\frac{c-1}{24}+h_{r,s}}}{\eta(q)^3}\bigg(q^{\frac{1}{2} l (r+s)+\frac{1}{2} m(r-s)}- \nn \\
& q^{\frac{1}{2} l (2 k-r-s+2)+\frac{1}{2} m (2 k-r+s+2)+k-r+1}\bigg),
\eea
and $\eta(q)$ is the Dedekind eta function.
The partition function is then given by 
\be 
Z(\tau) = \sum_{s=0}^{2k-1}\sum_{r=0}^{k}|\chi_{r, s}(\tau)|^2.
\label{eq:paraZ}
\ee
One may show that the theories for $k = 1,2,3$ correspond to Virasoro minimal models, but for $k \geq 4$ we have new RCFTs. In this paper, we include parafermionic theories with $k=4, 5, 6, 7$. The energy spectra of these models can be obtained by the $q$ expansion of Eq.~\eqref{eq:paraZ}.

\subsection{$\mathcal{N}=1$ superconformal minimal models}
The $\mathcal{N} = 1$ superconformal minimal models 
have central charge $c=\frac{3}{2}-\frac{12}{k(k+2)}$, with $k \geq 2$ an integer \cite{gils2013anyonic}. The scaling dimension of the primary field is
\be
h_{r, s}=\frac{[(k+2)r-ks]^2-4}{8k (k+2)}+\frac{1}{32} \left(1-(-1)^{r+s}\right).
\ee
where $1 \leq r \leq k-1$ and $1 \leq s \leq k+1$.
Fields with $r+s$ even have a conformal dimensions are given by:
\be
h^{'}_{r, s} = h_{r, s} + \frac{1}{2} + \delta_{r+s,2}.
\ee 

The characters and partition function for this case are much more envolved, and we refer the readers to Refs.~
\cite{matsuof1986superconformal, kiritsis1988character} for their explicit expressions.

\section{Preprocess procedure}
\label{appendix:para}
\paragraph{}
In this section, we discuss how to eliminate the non universal constants $E_0, E_1, L, v$. We refer to this procedure as preprocessing. We have the following relation,
\be
E=E_1L + E_0 + \frac{2\pi v}{L}\left(-\frac{c}{12}+h_L+h_R \right).
\ee
Let $ H_i$ be the $i$-th value of $h_L+h_R$, which is an integer. Notice that the lowest level, $H_0$, is zero for all CFTs investigated in this work. Defining $\{ X_0, X_1,...X_n \}$ as our non universal energies, we have 
\begin{equation}
X_i=E_1L + E_0 + \frac{2\pi v}{L}\left(-\frac{c}{12}+H_i \right).
\end{equation}
We then have the energy differences from the ground state,
\begin{equation}
\frac{2\pi v}{L}H_i=X_i-(E_1L + E_0)+\frac{2\pi v}{L}\frac{c}{12}.
\end{equation}
We then rescale the highest shifted energy to be 1. This gives a set of $n$ preprocessed energies, $\{ x_0, x_1,...x_n \}$, given by
\begin{equation}
x_i=\frac{X_i-(E_1L + E_0)+\frac{2\pi v}{L}\frac{c}{12}}{\frac{2\pi v}{L}H_n}=\frac{H_i}{H_n}.
\end{equation}
We see that $x_i$ is independent of $E_0, E_1, L, v$.

\bibliography{reference}

\end{document}